# Design and Evaluation of Self-Assembled Actin-Based Nano-Communication

Oussama Abderrahmane Dambri, Soumaya Cherkaoui and Biswadeep Chakraborty[†]
Department of Electrical and Computer Engineering, Université de Sherbrooke, Canada
Email: {abderrahmane.oussama.dambri, soumaya.cherkaoui}@usherbrooke.ca

[†]Department Electronics and Telecommunication Engineering, Jadavpur University, India
Email: {biswadeep96}@icloud.com

*Abstract*—The tremendous progress in nanotechnology over the last century, makes it possible to engineer tiny nanodevices, which they need a nano-communication network to interact. Two solutions are proposed in literature to create a nano-communication system, either by using the classical electromagnetic paradigm with Terahertz band, or using the bio-inspired molecular communication. However, Terahertz is suffering from molecular absorption and scattering losses at nano level, and the achievable throughput of molecular communication is very low. In this paper, we propose a new solution to establish a wired nano-communication. Self-assembled actin-based is a new method that takes advantage of actin filaments self-assembly to create a nano wire between a transmitter and a receiver, and use electrons as information carriers. VPython framework is used in this paper to perform stochastic simulations of the nano wire formation. The algorithms used for the simulations are presented. The stability of the constructed nano wire is analyzed, and the error probability is calculated. Self-assembled actin-based method promises a fast and stable nano-communication system with a very high achievable throughput.

## I. INTRODUCTION

In recent decades, a new generation of tiny nanomachines has emerged and quickly become fundamental tools in many fields. The limited capacity of these nanomachines has urged to construct nanonetwork systems that allow them to interact and cooperate with each other. Thus, the connection between the nanomachines helps overcome their energy and processing capacity limitations, and significantly increase their potential [1]. Nanonetworking is the paradigm that cope with communicating systems at nano level, by adapting the classical communication paradigms to meet the requirements of the nanosystems. Wired and wireless connections at nanoscale levels are very challenging, and to overcome this difficulty, new solutions are required.

The electromagnetic communication at nanoscale implies the use of Terahertz (THz) band, because nanomachines oscillate at THz frequency [2]. Despite the speed of THz waves and its huge frequency band, a lot of challenges are needed to be tackled in all communication layers in order to capture the peculiarities of THz. One of THz main challenges is the high path loss caused by the molecular absorption and the scattering losses [3]-[5]. Another challenge resides in the design of the transmitters and the receivers that can emit and detect the fast waves of THz [6]. The graphene with its physical properties offers new solutions to overcome the transceiver's design problems [7]. However, THz waves are highly absorbed by water molecules, which increases their vibrations and thus, increases the heat of the medium [8]. The increase of the heat can cause problems if THz is used inside the human body for medical applications, and further research is needed to investigate the safety of its use inside the body.

Another promising solution is proposed in literature to create a nano-communication system, which uses molecules to carry the information instead of electromagnetic waves. Inspired by nature, molecular communication uses the random diffusion of molecules in the medium to reach the receiver [9]-[11]. The main challenge of molecular communication based on diffusion is the Inter-Symbol Interference (ISI) caused by the remaining molecules in the medium from previous transmissions, which may interfere with future ones [12]-[15]. Other methods are also proposed in literature that use bacteria [16], [17], protein motors [18], [19], and calcium ions [20], [21], as propagation ways to carry the molecules. However, with all the proposed methods and techniques to enhance molecular communication performance, it still has a very limited achievable throughput and a higher delay.

In this paper, we propose a new solution to create a wired nano-communication system, based on the self-assembly of actin. Actin is a bi-globular protein that has the ability to construct a self-assembled filament, using specific enzymes that plays an important role in cell division, its movements and its skeletal maintenance [22]. The electrical properties of actin filaments allow it to play the role of a conductive wire in our proposed system [23]. However, the process of actin assembly is random and the direction of the filament is unpredictable. Therefore, a magnetic field is used in the system to guide the direction of the filament towards the receiver. This paper studies the formation of the actin filament that connects the transmitter to the receiver, by simulating a 3 dimensional system with spherical molecules randomly binding to each other. We used vPython to make 3D stochastic simulations in real time, and we conditioned the binding probability of molecules with the intensity of the magnetic field and with the enzyme concentration. The transmitter of the proposed system is a nano thread ZnO matrix with a piezoelectric propriety, which enables it to convert the mechanical energy into electrical energy [24]. We can use ultrasonic waves to vibrate the nano matrix, which will convert the vibration into electrons that will be conducted through the actin filament to the receiver, as shown in Fig. 1. The receiver of the system is constructed with a hybrid phospholipid/alkanethiol bilayers membrane, because of its insulating nature, in order to minimize the loss of electrons at the receiver [25]. When the sent electrons get inside the receiver, a bioluminescent reaction takes place and generates light, which can be sensed by a photodetector as the receiver response. The intensity of the emitted light at the receiver is dependent on the transmitted electrons intensity, which can be used as a method to modulate the information. In the next section, we provide a more detailed description of the system's design. We also present the 2 algorithms that we used as a framework to simulate the construction of the proposed nano wire. We then use the simulation results to empirically study the stability of the constructed nano wire, and we model the error probability of the system. We finally discuss the simulation results.

The rest of the paper is organized as follows. In section II, we highlight an in-depth description of the proposed system with more detailed explanations for the transmitter and the receiver design. In section III, we present and explain the 2 algorithms used in our framework to simulate the 3D system with Brownian motion of molecules to make the system more real. We derive the error probability in section IV, and we discuss the simulation results. Finally, we conclude the paper and we present future work in section V.

## II. SYSTEM DESIGN

To the best of our knowledge, the wired nano-communication system proposed in this paper, is the first attempt to use electrons as carriers of the information in a wired system at the nano level. The proposed system can be used safely inside the human body for medical applications. Its implementation is noninvasive due to its tiny size and electrons generation can be controlled from the outside by using ultrasonic waves. As shown in Fig. 1, the system contains a transmitter with ZnO matrix controlled by ultrasonic waves, assembled actin wire guided by a magnetic field and a receiver that generates bioluminescent light after absorbing the transmitted electrons. A photo-detector is added to the system to detect the emitted light and to play the role of a gateway.

### A. Transmitter

Wired communication at nano level needs nanogenerators to emit the electrons, and the piezoelectric property of the nano-materials is the best solution to do just that. The piezoelectric potential is created by the polarization of the ions in some solid materials such as crystals and ceramics, or biological matter as DNA and some proteins, when subjected to strain. The system proposed in this paper needs a transmitter that can generate electricity when subjected to mechanical force. In [24], the authors proposed an innovative approach that converts the mechanical energy into electricity at nano level by using a piezoelectric Zinc Oxide (ZnO) nanowire matrix. Using the unique coupling of semiconducting and piezoelectric dual ZnO properties and a Schottky barrier between the metal tip and the nanowire, the authors created an electric nanogenerator. The developed DC nanogenerator in [24] is driven by ultrasonic waves in a bio-fluid. These ultrasonic waves are used in our proposed system to send the information, which will be converted by the nanogenerator into electricity and transmitted through the constructed nanowire. The amplitude of the waves and their frequency can be used to modulate the information.

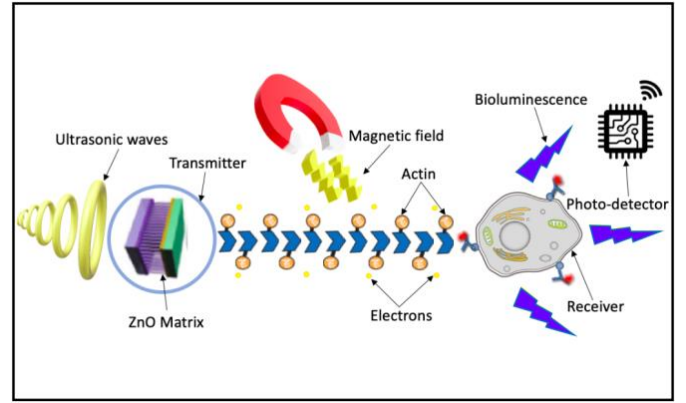

Fig. 1. System design, where ZnO matrix is the transmitter of the electrons by transforming the ultrasonic waves into electricity. The self-assembled actin guided by the magnetic field, connects the transmitter to the receiver and transports the electrons. When the receiver absorbs the electrons, it emits a bioluminescent light. The emitted light can be detected by a photo-detector, which can play the role of a gateway. The intensity of the detected bioluminescent light is proportional to the transmitted electrons, which are controlled by ultrasonic waves.

### B. Self-Assembled Actin Wire

All cells of all living creatures contain a complex of interlinking filaments called "cytoskeleton", which has essential roles in movement, shape and division of cells [22]. The cytoskeleton contains three types of these filaments: microfilaments, microtubules and intermediate filaments. This latter is constructed by specific proteins such as keratin, desmin, lamin and vimentin, the microtubule is constructed with Tubulin and the microfilament is formed with actin. Actin is one of the most abundant proteins, comprising 10% of the total proteins in muscle cells and around 5% in the other cells [26]. Compared to microtubules, microfilaments constructed by actin are more flexible, and their polymerization process is easily controllable, which justify our choice to use actin as the best candidate to construct our proposed nanowire. A study in [27] proposed the design of conductive actin-based metallic nanowire, which exhibited a very high electrical conductivity. However, the polymerization process of the Globular actin (G-actin) to form a Filament actin (F-actin) is random, because of the thermal fluctuations in the medium. Guiding the assembled nanowire to the receiver is one of the main challenges to construct a wired nano-communication in a randomly diffusive medium. Two solutions are proposed in literature to guide the actin assembly, either by using electric or magnetic field. In [28], the authors used AC electric field to align F-actin in a desired direction. Using AC electric field inside the human body, however, is dangerous, therefore, the study in [29] proposed the use of a bar magnet with low-intensity magnetic field (22 mT). The authors proved that the majority of the F-actin got permanently aligned towards the magnetic lines. The study concluded that magnetic field can be used to permanently orient and guide the alignment of F-actin towards a desired direction. In our proposed system, we use the magnetic field to guide the actin assembly towards the receiver.

*C. Receiver*

When the nanowire reaches the destination, the last assembled G-actin binds with one of G-actin proteins already anchored at the membrane, which facilitates the binding between the nanowire and the receiver. The membrane is constructed with a hybrid phospholipid/alkanethiol bilayers, due to its insulating nature. The G-actin is anchored at the receiver surface using electrodes, which helps the electrons to pass through the receiver. The transmitted electrons in our proposed system are used to generate bioluminescent light inside the receiver. Bioluminescence is a chemical emission of light by living organisms using light-emitting molecules and enzymes. The most famous light-emitting molecule is "Luciferin" and its enzyme "Luciferase", which catalyzes the oxidation of Luciferin, generating by that light energy. Another photoprotein called "Aequorin" constructed by the jellyfish *Aequorea Victoria*, which can be found in North America and Pacific Ocean [30]. Aequorin is activated in the presence of $Ca^{2+}$ and generates blue light as described in Fig. 2 [31]:

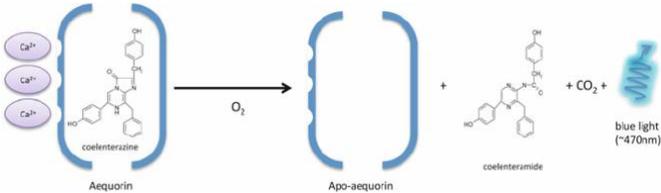

Fig. 2. Bioluminescent reaction that uses Aequorin, which in the presence of $Ca^{2+}$ ions, it generates blue light [31].

As shown in Fig. 3, we propose a receiver design containing a Smooth Endoplasmic Reticulum (SER), which works as a $Ca^{2+}$ ions storage in cells [32]. When SER is excited by the transmitted electrons, it secretes $Ca^{2+}$ ions inside the receiver. In the presence of $Ca^{2+}$ ions, Aequorin becomes activated and emits blue light, which it will be sensed by the photo-detector. When the electron transmission stops, the SER absorbs all the $Ca^{2+}$ ions inside the receiver as a sponge and Aequorin stops the blue light emission.

## III. SIMULATION FRAMEWORK

In this study, we simulate only the nanowire formation, the transmitter and the receiver simulations will be considered in future work. We used GlowScript VPython to simulate a real-time 3D animation of the nanowire assembly using random diffusive molecules. GlowScript is an easy-to-use, powerful environment for creating 3D simulations and publishing them on the web where the programs are written using vPython. RapydScript is a pre-compiler for JavaScript written front-end in Python, with high performance and interoperability with external JavaScript libraries. We used 2 algorithms in our framework, which will be described in the next subsections.

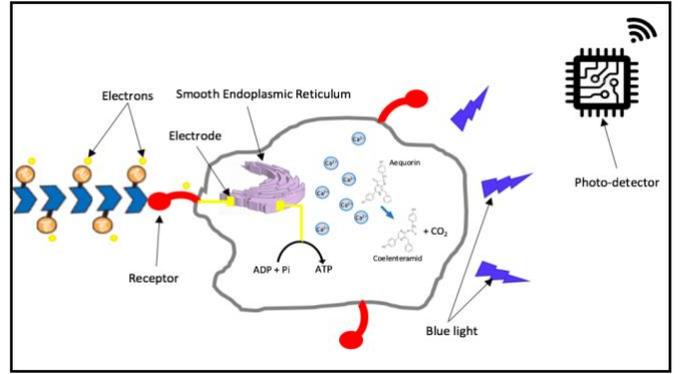

Fig. 3. Receiver design that uses electrons to generate bioluminescent blue light.

*A. Collision Between the Molecules*

The algorithm contains two strategies, Periodic Interference Test (PIT) and a Predicted Instant of Collision (e r). In (PIT), at every frame in the simulation, there occurs a check to detect if there is any collision between any two molecules. This check is done by testing if the molecules are approaching each other, and whether they interfere spatially. If a collision is detected, the new velocities after the shock are computed using the equations for conservation of momentum and energy; and the molecules move along the new path with the new velocity. If there is no collision in that frame, the simulation moves along to the next frame. There occurs no check between two successive frames to detect a possible collision. In PIC, on the other hand, the time to the earliest collision is calculated. If the time is not earlier than the next frame instant, the simulation continues until the frame in which collision occurs. When collision occurs, the velocities of the bodies after the shock are calculated. The bodies then move along the new path with the new velocity. This method pre-calculates the exact time of collisions, and does not allow the molecules to interfere at all. The PIC approach models a collision perfectly, at the cost of some extra computation, compared to the PIT approach, which is a lazy collision detection strategy.

---

**Algorithm 1**: Periodic Interference Test

**Procedure** PITC (Molecule A, Molecule B):
    At every frame instant t:
    Test if the two molecules are approaching each other.
    **if** YES **then**
        Collision detected.
        Compute new velocities.
        T:= time of the collision between a and B.
        **if** T < next frame instant: **then**
            Move frame to intermediate instant T.
            Calculate velocities after collision.
        **end if**
    **else**
        No collision detected.
    **end if**
    Goto next frame instant.
    **else**
        Goto next frame instant.
    **end if**
**end procedure**

## B. Nanowire Formation

When a molecule is struck by one of the molecules in the nanowire, the molecule is attached to the nanowire, and thus increasing its length. We maintain an array of the molecules number in the nanowire. The sticking of the molecules is done if the following conditions are met:

- One of the molecules having a collision is in the array of the molecules in the nanowire.

- The center of the molecule which is striking lies within a specified angle range from the center of the last attached molecule.

- The center of the next molecule being attached should be farther away from the transmitter and closer to the receiver, so that the nanowire is moving towards the receiver and not towards any random path.

To make the molecules stick, we make their momentum zero and we do not update their position with time.

**Algorithm 2**: Nanowire Formation

```
molecule_array = [Transmitter]
if collision.detected = True then
  if collision occurs with last (molecule_array) && last(molecule_array)!= Receiver then
        M←molecule hitting the wire
        if Magnetic Field = 0 then
          Randomly attach M to last(molecule_array) in any direction
          molecule_array.append(M)
        end if
        if Magnetic Field != 0 then
          if M.center.X > max(last(molecule_array).position) then
            Z = function(magnetic field)
            if M.center.Z < max(Z) then
                M.momentum = 0 % Stick the two molecules together
                molecule_array.append(M)
            end if
          end if
        end if
  end if
end if
```

The direction algorithm regulates the direction of the nanowire formation. It is controlled by the magnetic field. If the magnetic field is zero, then there is no direction, and the nanowire can form in any direction. As the magnetic field increases, the nanowire gets more and more aligned to the straight line joining the transmitter and the receiver.

## IV. EVALUATION AND SIMULATION RESULTS

The simulated system is a bounded environment containing two big spheres, which represent the transmitter and the receiver, and smaller spheres representing the assembled actin, as shown in Fig. 4. The simulation is an oversimplification of what really happens in nature, to facilitate the calculations. Actin assembly is controlled by specific enzymes in each step of the polymerization and depolymerization processes. In our simulation, molecules diffuse randomly in the medium and when a molecule collides with the transmitter, the assembly starts and continue until the last assembled molecule reaches the receiver. We add a magnetic field as a parameter to guide the direction of the molecules assembly towards the receiver. The direction is proportional to the magnetic field intensity. If the intensity is very small, the direction is more random, otherwise it is aligned. We also include an enzyme parameter that influence the binding probability of molecules to make the simulation more real. The increase in the enzyme concentration increases the probability that two molecules bind together. Our simulator gives a graph that plots the position of the last assembled molecule in real time as shown in Fig. 5, which allows us to study the speed of the nanowire formation.

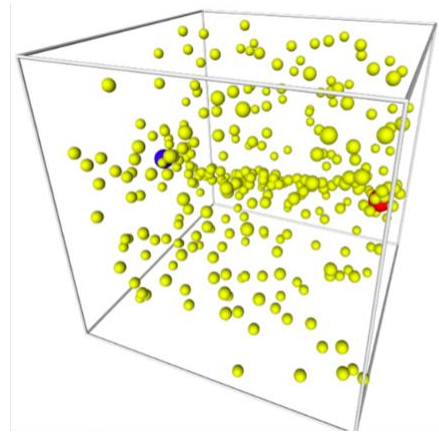

Fig. 4. Actin-based nanowire formation represented by small yellow spheres, which link the transmitter (blue sphere) to the receiver (red sphere).

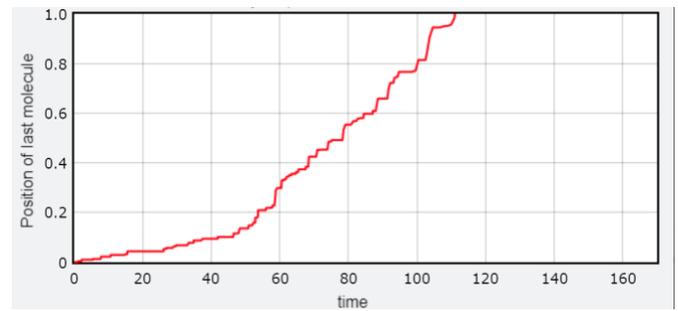

Fig. 5. Position of the last assembled molecule as a function of time representing the speed of the nanowire formation.

## A. Parameters Variation

We changed the magnetic field intensity and the enzyme concentration to study their influence on the nanowire formation. In Fig. 6, the graph shows the position of the last assembled molecule (μm) as a function of time (min) with two different values of the magnetic field intensity. We can notice that when the value is high (M=20), the speed of the nanowire formation increases compared to its speed with the smallest value (M=10). This result can be explained by the fact that when the magnetic field intensity is low, the direction of the nanowire formation becomes more and more random, and takes more time to reach the receiver.

The same result is obtained when we changed the enzyme concentration in Fig. 7. High concentrations of the enzymes increase the speed of nanowire formation (C=1, C= 0.8), which is logical because the enzymes control the molecules binding. The high concentration of enzymes in the medium increases the probability of molecules binding, which makes the nanowire formation faster. The parameter values used in the simulations are still abstract, but an equivalence with real scenarios in nature is in progress for future work.

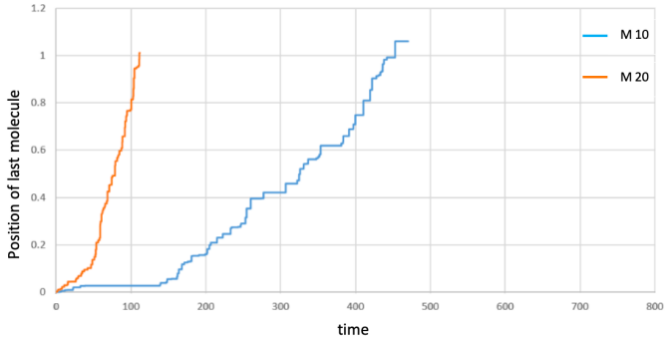

Fig. 6. Position of the last assembled molecule as a function of time for two different magnetic field intensity values (M=10, M=20).

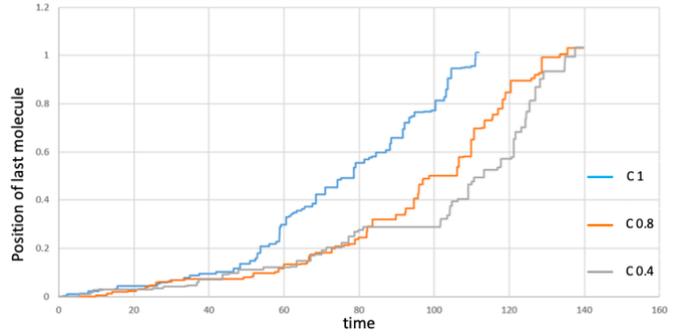

Fig. 7. Position of the last assembled molecule as a function of time for three different enzyme concentration values (C=1, C=0.8, C=0.4).

*B. Stability*

The parameters studied above affect the nanowire stability, the increase in magnetic field intensity and enzyme concentration increases the stability of the nanowire. The distance between the transmitter and the receiver can also affect the stability of the nanowire, because actin filaments show less stability when they become too long. Therefore, we can empirically study the stability of the nanowire by using the following equation:

$$Stability = k \frac{E \times M}{L} \quad (1)$$

Where *E* is the enzyme concentration, *M* is the intensity of the magnetic field and *L* is the length of the nanowire.

The simulation result of the equation (1) is shown in Fig. 8. We can see that the more the length of the nanowire is increased, the less its stability. We can also notice that the increase in the enzyme concentration and the magnetic intensity increases the stability of the nanowire.

*C. Error Probability*

The error probability of the proposed communication system can be written as:

$$P_e = p_0 P_{\epsilon 0} + p_1 P_{\epsilon 1}, \quad (2)$$

Where $\boldsymbol{p_0} = \boldsymbol{p_1} = 0.5$ are the *a priori* probabilities, $\boldsymbol{P_{\epsilon 0}}$ is the probability of error for a bit '0' and $\boldsymbol{P_{\epsilon 1}}$ is the probability of error for a bit '1'. We used the normal distribution to model the noise, in our case, the noise is generated from the constant assembly and disassembly of the nanowire. However, the assembly is more frequent than the disassembly in the process of the nanowire formation, though, the wrong bit is more probably 0 than 1. So, the noise distribution of the bit 0 will remain Gaussian, and the noise distribution of the bit 1 will be skewed left. We know that

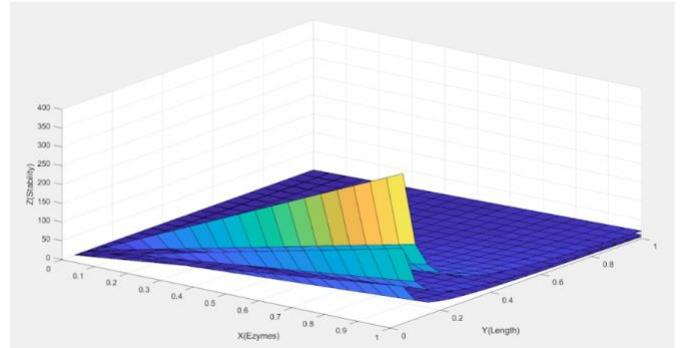

Fig. 8. Enzymes, length and magnetic field effects on the nanowire stability.

as the standard deviation increases, the probability distribution becomes flatter, and thus the probability of correctly receiving a bit decreases. Therefore, we can model the standard deviation as the inverse stability defined in (1) as $\boldsymbol{SD = k' \frac{L}{EM}}$ and we can write:

Where *A* is the skewness coefficient of the distribution. By

$$P_{\epsilon 0} = \frac{1}{\sqrt{2\pi}} e^{\frac{-x^2}{2}}, \quad (3)$$

$$P_{\epsilon 1} = \frac{1}{\sqrt{2\pi}} e^{\frac{-(x-A)^2}{2}} \operatorname{erfc}\left(\frac{x-A}{\sqrt{2}}\right), \quad (4)$$

$$P_e = \frac{1}{2\sqrt{2\pi}} \left( e^{\frac{-(x-A)^2}{2}} \operatorname{erfc}\left(\frac{x-A}{\sqrt{2}}\right) + e^{\frac{-x^2}{2}} \right). \quad (5)$$

substituting (3) and (4) in (2), we write the error probability as:

Since the distribution is negatively skewed, *A* is also negative. For a skew normal distribution for which the scale factor is 1, the variance is given by $\boldsymbol{1 - \frac{2\delta^2}{\pi}}$, where $\boldsymbol{\delta = \frac{A}{\sqrt{1+A^2}}}$.

V. CONCLUSION AND FUTURE WORK

Self-assembled actin-based is, to the best of our knowledge, the first wired nano-communication system, which uses electrons as carriers of the information. The system uses the self-assembly ability of actin proteins to create a nanowire that connects the transmitter to the receiver at nano level. This paper explains the proposed system design which contains a transmitter that converts ultrasonic waves into electricity, a nanowire guided by a magnetic field, and a receiver that uses the electrons to generate bioluminescent light. We used GlowScript VPython to simulate the nanowire formation, and we presented the algorithms used in our framework. We also studied the stability of the nanowire and calculated the error probability. The proposed method promises a fast and stable nano-communication system with a very high achievable throughput.

In future work, we will consider simulating the transmitter and the receiver of the proposed system, and provide an analytical model for the nanowire formation. We will also extend the system design to become more complex with clusters, where the nodes of each cluster connect to each other using the nanowire, and each cluster uses the photo-detector as a gateway.